\documentstyle[preprint,tighten,aps,psfig]{revtex}

\begin{document}
\title{ Chiral multiplets of excited mesons.}
\author{ L. Ya. Glozman}
\address{  Institute for Theoretical
Physics, University of Graz, Universit\"atsplatz 5, A-8010
Graz, Austria\footnote{e-mail: leonid.glozman@uni-graz.at}}
\maketitle

\begin{abstract} 
It is shown that experimental meson states with spins
J=0,1,2,3 in the energy range 1.9 - 2.4 GeV obtained
in a recent partial wave analysis of proton-antiproton
annihilation at LEAR remarkably confirm all predictions
of chiral symmetry restoration. Classification of excited
$\bar q q$ mesons according to the representations of
chiral $U(2)_L \times U(2)_R$ group is performed. There
are two important predictions of chiral symmetry restoration 
in highly excited mesons: (i) physical states must fill out
approximately degenerate parity-chiral multiplets; (ii)
some of the physical states with the given $I,J^{PC}$ are
 members of one parity-chiral multiplet, while the
other states with the same $I,J^{PC}$ are  members
of the other parity-chiral multiplet. For example, while
some of the excited $\rho(1,1^{--})$ states are systematically
degenerate with $a_1(1,1^{++})$ states forming (0,1)+(1,0)
chiral multiplets, the other excited $\rho(1,1^{--})$ states 
are degenerate with $h_1(0,1^{+-})$ states ((1/2,1/2) chiral
multiplets). Hence, one of the predictions of chiral symmetry
restoration is that the combined amount of  $a_1(1,1^{++})$ and
$h_1(0,1^{+-})$  states must coincide with the amount of
$\rho(1,1^{--})$ states in the chirally restored regime. It
is shown that the same rule applies (and experimentally
confirmed) to many
other meson states.
\end{abstract}

\bigskip
\bigskip

\section{Introduction}
A recent partial wave analysis \cite{BUGG1,BUGG2,BUGG3,BUGG4}
of  $\bar p p$ annihilation at LEAR in the energy range
1.9 - 2.4 GeV has revealed a lot of new meson states in this
mass region. Very many of them do not fit into  the traditional
potential description which is based on the $^{2S+1}L_J$
classification scheme. For instance, 
they cannot be accommodated by the
constituent quark model \cite{GI}. Some of the particular
examples are discussed in refs. \cite{G1,G2}. If these experimental
results are correct, then it means that a "harmony" of the
potential description is certainly not adequate for highly
excited states (where the valence quarks should be expected to
be ultrarelativistic) and some other physical picture and classification
scheme must be looked for.\\

\begin{figure}
\centerline{
\psfig{file=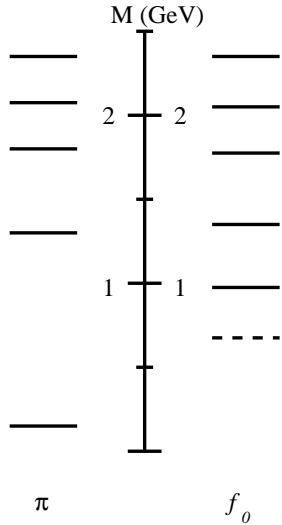,angle=-90,width=0.6\textwidth}
}
\caption{Pion and $n \bar n=\frac{\bar u u + \bar d d}{\sqrt 2}$ $f_0$ spectra. 
Since  these $f_0$ states are obtained in $p \bar p$ and   they
decay predominantly  into $\pi \pi$ channel, they are considered  
 as  $n \bar n$ states.}
\end{figure}

In ref. \cite{G3} it has been suggested that spontaneously broken
chiral symmetry of QCD must be restored in the upper part
of hadron spectra, which is evident by the (almost) systematical
parity doubling of $N$ and $\Delta$ states at $M \geq 1.7$ GeV. This
idea has been substantiated   on the basis
of the operator product expansion in QCD and analyticity of the 
two-point function (dispersion relation) \cite{CG1,CG2}. This
effective chiral symmetry restoration has been referred to as
chiral symmetry restoration of the second kind \cite{G4} in order
to distinguish this phenomenon from the other phenomena of chiral
symmetry restoration in the vacuum state at high temperature and/or
density. The essence of the present phenomenon is that the quark 
condensates which break chiral symmetry in the vacuum state (and
hence in the low-lying excitations over vacuum) become simply
irrelevant (unimportant) for the physics of the highly excited states
and the physics here is such as if there were no chiral symmetry
breaking in the vacuum. The valence quarks simply decouple from
the quark condensates and consequently the notion of the constituent
quarks with dynamical mass induced by chiral symmetry breaking
becomes irrelevant in highly excited hadrons \cite{G3,swanson}.
Instead,
the string picture with quarks of  definite chirality at the
end points of the string should be invoked \cite{G2}.
In  recent lattice calculations DeGrand has demonstrated that
indeed in the highly excited mesons valence quarks decouple
from the low-lying eigenmodes of the Dirac operator (which determine
the quark condensate via  Banks-Casher relation) and hence decouple
from the quark condensate of the QCD vacuum \cite{degrand}.\\

If it is true that chiral symmetry is effectively restored, then
classification of hadrons should be performed according to the
chiral group of QCD. This group strongly constrains the amount of 
certain mesons and their relative energies. In ref. \cite{G1}
we have classified all excited mesons with spin $J=0$ according
to the chiral group and have shown that indeed the pattern
of these mesons above 1.7 GeV region certainly 
favours   chiral symmetry restoration of the second kind.
This is illustrated in Fig. 1 where $\pi(I=1,0^{-+})$ and
$\bar q q$   $f_0(I=0,0^{++})$ states  \cite{G1} are shown (which must be
chiral partners in the chiral symmetry restored regime).\\

The symmetry restoration implies that it must
be seen also in all other $\bar q q$ mesonic states. The purpose
of this paper is to classify observed $\bar q q$ mesons according
to chiral representations and  demonstrate that the experimental
patterns of $J=1,2,3$ mesons do confirm this phenomenon. Experimental
data on higher spin states are scarce and it is not yet possible
to provide any systematic description though predictions can
be made.

\section{Classification of the $\bar q q$ mesons according to
the chiral group.}

Mesons reported in ref. \cite{BUGG1,BUGG2,BUGG3,BUGG4} are obtained
in $\bar p p$ annihilations,
 hence according to OZI rule we have to expect them
to be $\bar q q$ states with $u$ and $d$ valence quark content. Hence
we will    consider  

\begin{equation}
U(2)_L \times U(2)_R = SU(2)_L\times SU(2)_R \times U(1)_V\times U(1)_A, 
\label{sym}
\end{equation}

\noindent
the full chiral group of the QCD Lagrangian. In the following
 chiral symmetry  will refer to specifically the $SU(2)_L\times SU(2)_R$
symmetry, which is  perfect due to very small
masses of $u,d$ quarks as compared to the only dimensional
parameter of QCD, $\Lambda_{QCD}$, and the typical hadronic scale
of 1 GeV.\\

$SU(2)_L\times SU(2)_R$ is a symmetry with respect to two
independent rotations of the left-handed and right-handed quarks in the
isospin space. Hence the irreducible representations of 
this group can be specified by the isospins of the left and
right quarks, $(I_L,I_R)$. The total isospin of the state
can be obtained from the left and right isospins according to
the standard angular momentum addition rules

\begin{equation}
I= |I_L-I_R|, ... , I_L+I_R.
\label{isospin}
\end{equation}

All hadronic states are characterised by a definite parity. However,
not all  irreducible representations of the chiral group are
invariant under parity. Indeed,  parity transforms the left
quarks into the right ones and vice versa. Hence while representations
with $I_L=I_R$ are invariant under parity (i.e. under parity
operation every state in the representation transforms into the
state of opposite parity within the same representation), this
is not true for the case $I_L \neq I_R$. In the latter case 
parity transforms every state in the representation $(I_L,I_R)$
into the state in the representation $(I_R,I_L)$. We can
construct definite parity states only combining basis vectors from both
these irreducible representations. Hence it is only the direct sum of these
two representations
 
\begin{equation}
(I_L,I_R) \oplus (I_R, I_L), ~~~~~~ I_L \neq I_R.
\label{mult}
\end{equation}

\noindent
 that is invariant under parity. This reducible representation of the
chiral group is an irreducible representation of the larger
group, parity-chiral group \cite{CG2}.\\

 When we consider mesons
of isospin $I=0,1$, only three independent irreducible representations
of the parity-chiral group exist.\\

{\bf (i)~~~ (0,0).} Mesons in this representation must have isospin
$I=0$. At the same time $I_R=I_L=0$. This can be achieved when 
either there are no valence quarks in the meson\footnote{Hence glueballs must
be classified according to this representation \cite{G5}; with
no quark content this representation contains the state of only
one parity.}, or both
valence quark and antiquark are right or  left.
If we denote $R=(u_R,d_R)$  and $L=(u_L,d_L)$, then the basis
states of both parities can be written as

\begin{equation}
|(0,0); \pm; J \rangle = \frac{1}{\sqrt 2} (\bar R R \pm \bar L L)_J.
\label{00}
\end{equation} 

\noindent
Note that such a system can have spin $J \geq 1$. Indeed, valence
quark and antiquark in the state (\ref{00}) have definite
helicities, because generically helicity = +chirality for quarks and
helicity = -chirality for antiquarks. Hence the total spin projection
of the quark-antiquark system  onto the momentum direction of the quark
is $ \pm 1$. The parity transformation property of the quark-antiquark state
is then regulated by the total spin of the system \cite{LANDAU}

\begin{equation}
\hat P |(0,0); \pm; J \rangle = \pm (-1)^J  |(0,0); \pm; J \rangle.
\label{P00}
\end{equation}

{\bf (ii)~~~ (1/2,1/2).} In this case the quark must be right
and the antiquark must be left, and vice versa. These representations combine
states with I=0 and I=1, which must be of opposite parity.
The basis states
within the two distinct representations of this type are

\begin{equation}
|(1/2,1/2); +;I=0; J \rangle = \frac{1}{\sqrt 2} (\bar R L + \bar L R)_J,
\label{I0+1}
\end{equation} 

\begin{equation}
|(1/2,1/2); -;I=1; J \rangle = \frac{1}{\sqrt 2} (\bar R \vec \tau L -
\bar L \vec \tau R)_J,
\label{I0+2}
\end{equation} 

\noindent
and

\begin{equation}
|(1/2,1/2); -;I=0; J \rangle = \frac{1}{\sqrt 2} (\bar R L - \bar L R)_J,
\label{I0-1}
\end{equation}

\begin{equation}
|(1/2,1/2); +;I=1; J \rangle = \frac{1}{\sqrt 2} (\bar R \vec \tau L +
\bar L \vec \tau R)_J.
\label{I0-2}
\end{equation}

\noindent
In these expressions $\vec \tau$ are isospin Pauli matrices.
The parity of  every state in the representation is determined as

\begin{equation}
\hat P |(1/2,1/2); \pm; I; J \rangle = \pm (-1)^J |(1/2,1/2); 
\pm; I; J \rangle.
\label{P12}
\end{equation}

Note that the two distinct (1/2,1/2) irreducible representations
of $SU(2)_L \times SU(2)_R$ form one irreducible representation
of $U(2)_L \times U(2)_R$.

{\bf (iii)~~~ (0,1)$\oplus$(1,0).} The total isospin is 1 and
the quark and  antiquark must both be right or  left.
 This representation
is possible only for $J \geq 1$. The basis states are

\begin{equation}
|(0,1)+(1,0); \pm; J \rangle = \frac{1}{\sqrt 2} (\bar R \vec \tau R 
\pm \bar L  \vec \tau L)_J
\label{10}
\end{equation} 

\noindent
with  parities

\begin{equation}
\hat P |(0,1)+(1,0); \pm; J \rangle = \pm (-1)^J 
 |(0,1)+(1,0); \pm; J \rangle.
\label{P10}
\end{equation} 
\\

In the chirally restored regime the physical states must fill
out completely some or all of these representations.
We have to stress that usual quantum numbers $I,J^{PC}$ are not
enough to specify the chiral representation. It happens that some of
the physical particles with the given $I,J^{PC}$ belong to
one chiral representation (multiplet), while the other particles with
the same $I,J^{PC}$ belong to the other multiplet. Classification
of the particles according to $I,J^{PC}$ is simply not complete
in the chirally restored regime. This property will have very
important implications as far as the amount of the states
with  the given  $I,J^{PC}$ is
concerned.\\

In order to make this point clear, we will discuss some
of the examples. Consider $\rho(1,1^{--})$ mesons. Particles
of this kind can be created from the vacuum by the vector
current, $\bar \psi \gamma^\mu \vec \tau \psi$. Its chiral
partner is the axial vector current, 
$\bar \psi \gamma^\mu \gamma^5 \vec \tau \psi$, which creates
from the vacuum the axial vector mesons, $a_1(1,1^{++})$ . Both
these currents belong to the representation (0,1)+(1,0) and
have the  right-right $\pm$ left-left quark content. 
Clearly, in
the chirally restored regime the mesons created by these currents
must be degenerate level by level and fill out the (0,1)+(1,0)
representations. Hence, naively the amount of $\rho$ and $a_1$
mesons high in the spectrum should be equal.
This is not correct, however.   $\rho$-Mesons can be also created
from the vacuum by  other type(s) of  current(s),
$ \bar \psi \sigma^{0i} \vec \tau \psi$ (or by 
$\bar \psi \partial^\mu \vec \tau \psi$). These interpolators belong
to the (1/2,1/2) representation and have the 
left-right $\pm$ right-left quark content.
In the regime where chiral symmetry is strongly broken (as
in the low-lying states) the physical states are mixtures
of different representations. Hence these low-lying states are
well coupled to both (0,1)+(1,0) and (1/2,1/2) interpolators.
However, when chiral symmetry is  (approximately)
restored, then each physical
state must be strongly dominated by the given representation
and hence will  couple only to the interpolator which belongs
to the same representation.
This means that 
$\rho$-mesons created
by two distinct currents in the chirally restored regime represent
physically different particles.  The chiral partner of the
$ \bar \psi \sigma^{0i} \vec \tau \psi$ (or 
$\bar \psi \partial^\mu \vec \tau \psi$) current is
$ \varepsilon^{ijk} \bar \psi \sigma^{jk}  \psi$ (
$\bar \psi \gamma^5 \partial^\mu   \psi$, respectively)
\footnote{Chiral transformation properties of some 
interpolators can be found in ref. \cite{CJ}.}. The
latter interpolators create from the vacuum $h_1(0,1^{+-})$
states. Hence in the chirally restored regime, some of the
$\rho$-mesons must be degenerate with the $a_1$ mesons ((0,1)+(1,0)
multiplets), but the others
- with the $h_1$ mesons ((1/2,1/2) multiplets)\footnote{
Those $\rho(1,1^{--})$ and $\omega(0,1^{--})$ mesons which
belong to (1/2,1/2) cannot be seen in $e^+e^- \rightarrow hadrons$.}. 
Consequently, high in the spectra the combined amount
of $a_1$ and $h_1$ mesons must coincide with the amount of
$\rho$-mesons. This is a highly nontrivial prediction of chiral
symmetry.\\

Actually it is a very typical situation. Consider 
  $f_2(0,2^{++})$ mesons as another example. They can be interpolated
by the tensor field $\bar \psi \gamma^\mu \partial^\nu \psi$
(properly symmetrised, of course), which belongs to the (0,0)
representation. Their chiral partners are $\omega_2(0,2^{--})$
mesons, which are created by the 
$\bar \psi \gamma^5\gamma^\mu \partial^\nu \psi$ interpolator.
On the other hand  $f_2(0,2^{++})$ mesons can also be created from
the vacuum by the $\bar \psi \partial^\mu \partial^\nu \psi$
type of interpolator, which belongs to the (1/2,1/2) representation.
Its chiral partner is 
$\bar  \psi \gamma^5\partial^\mu \partial^\nu \vec \tau \psi$,
which creates $\pi_2(1,2^{-+})$ mesons. Hence in the chirally
restored regime we have to expect $\omega_2(0,2^{--})$ mesons
to be degenerate systematically with some of the $f_2(0,2^{++})$ mesons
((0,0) representations) while $\pi_2(1,2^{-+})$ mesons must
be degenerate with {\it other} $f_2(0,2^{++})$ mesons (forming
(1/2,1/2) multiplets). Hence the total number of $\omega_2(0,2^{--})$
and $\pi_2(1,2^{-+})$ mesons in the chirally restored regime
must coincide with the amount of $f_2(0,2^{++})$ mesons.\\

These examples can be generalized to mesons of any spin $J \geq 1$.
Those interpolators which contain only derivatives
$\bar \psi \partial^\mu \partial^\nu ...\psi$ ( 
$\bar  \psi \vec \tau \partial^\mu \partial^\nu ...\psi$)
have quantum numbers $I=0,P=(-1)^J,C=(-1)^J$ ($I=1,P=(-1)^J,C=(-1)^J$)
and transform as (1/2,1/2).
Their chiral partners are 
$\bar \psi \vec \tau  \gamma^5 \partial^\mu \partial^\nu ...\psi$
( $\bar  \psi \gamma^5\partial^\mu \partial^\nu ...\psi$, respectively)
with $I=1,P=(-1)^{J+1},C=(-1)^J$ ($I=0,P=(-1)^{J+1},C=(-1)^J$,
respectively). However, interpolators
with the same $I,J^{PC}$ can be also obtained with one $\gamma^\eta$
matrix instead    one of the  derivatives, $\partial^\eta$:
$\bar \psi \partial^\mu \partial^\nu ...\gamma^\eta...   \psi$ ( 
$\bar  \psi \vec \tau \partial^\mu \partial^\nu ... \gamma^\eta ...\psi$).
These
latter interpolators  belong  to (0,0)  ((0,1)+(1,0)) representation. 
Their chiral partners are
$\bar \psi \gamma^5 \partial^\mu \partial^\nu ...\gamma^\eta...   \psi$ ( 
$\bar  \psi \vec \tau  \gamma^5 \partial^\mu \partial^\nu ... 
\gamma^\eta ...\psi$) which have 
$I=0,P=(-1)^{J+1},C=(-1)^{J+1}$ ($I=1,P=(-1)^{J+1},C=(-1)^{J+1}$).
Hence in the chirally restored regime
the physical states created by these  different types of interpolators
 will belong to different
representations and will be distinct particles
while having the same $I,J^{PC}$. One needs to indicate
chiral representation in addition to  usual quantum numbers
$I,J^{PC}$ in order to uniquely  specify physical states in the
chirally restored regime.\\

In the next section we will present experimental data and show
that all these predictions of chiral symmetry are indeed verified.

\section{Chiral multiplets of $J=1,2,3$ mesons.}

A detailed analysis of all chiral multiplets with $J=0$ has
been performed in ref. \cite{G1} and we do not repeat it here.
In Table 1 we present all mesons with $J=1,2,3$ in the energy range
1.9-2.4 GeV obtained in refs. \cite{BUGG1,BUGG2,BUGG3,BUGG4}.
So it is important to see whether these experimental results
fall into chiral multiplets  and whether the data set
is complete. Below we will show that indeed these data remarkably
confirm the predictions of chiral symmetry restoration. We will
start with chiral multiplets having $J=2$ because the data seems to be
 indeed complete for the $J=2$ states.\\

\begin{center}{\bf (0,0)}\\

{$\omega_2(0,2^{--})~~~~~~~~~~~~~~~~~~~~~~~f_2(0,2^{++})$}\\
\medskip
{$1975 \pm 20   ~~~~~~~~~~~~~~~~~~~~~~~~1934 \pm 20$}\\
{$2195 \pm 30   ~~~~~~~~~~~~~~~~~~~~~~~~2240 \pm 15$}\\

\bigskip
{\bf (1/2,1/2)}\\

{$\pi_2(1,2^{-+})~~~~~~~~~~~~~~~~~~~~~~~f_2(0,2^{++})$}\\
\medskip
{$2005 \pm 15   ~~~~~~~~~~~~~~~~~~~~~~~~2001 \pm 10$}\\
{$2245 \pm 60   ~~~~~~~~~~~~~~~~~~~~~~~~2293 \pm 13$}\\

\bigskip
{\bf (1/2,1/2)}\\

{$a_2(1,2^{++})~~~~~~~~~~~~~~~~~~~~~~~ \eta_2(0,2^{-+})$}\\
\medskip
{ $2030 \pm 20  ~~~~~~~~~~~~~~~~~~~~~~~~2030 ~\pm ~?$}\\
{ $2255 \pm 20 ~~~~~~~~~~~~~~~~~~~~~~~~2267 \pm 14$}\\

\bigskip
{\bf (0,1)+(1,0)}\\

{$a_2(1,2^{++})~~~~~~~~~~~~~~~~~~~~~~~\rho_2(1,2^{--})$}\\
\medskip
{ $1950^{+30}_{-70}~~~~~~~~~~~~~~~~~~~~~~~~~~~1940 \pm 40$}\\
{ $2175 \pm 40  ~~~~~~~~~~~~~~~~~~~~~~~~2225 \pm 35$}\\
\end{center}

We see systematic patterns of chiral symmetry restoration. In
particular, the amount of $f_2(0,2^{++})$ mesons coincides with
the combined amount of $\omega_2(0,2^{--})$ and $\pi_2(1,2^{-+})$
states. Similarly,  number of $a_2(1,2^{++})$ states is
the same as number of $\eta_2(0,2^{-+})$ and $\rho_2(1,2^{--})$
together. All chiral multiplets are complete. While masses of
some of the states will be definitely corrected in the future experiments,
if  new  states might be discovered in this energy region in other
types of experiments, they
should be either $\bar s s$ states or glueballs.\\

Consider now the $J=1$ multiplets\footnote{The state $\rho (1,1^{--})$
mentioned in the Table at $2110 \pm 35$ is given below
at the mass 2150, because it is this value for this state
which is given in PDG \cite{PDG}. The $\rho (1,1^{--})$ state at 1900, used
in the classification, is seen in $e^+e^-$ (see PDG) and not seen in
$\bar p p$.}.

\begin{center} {\bf (0,0)} \\

{$\omega(0,1^{--})~~~~~~~~~~~~~~~~~~~~~~~f_1(0,1^{++})$}\\
\medskip
{$~~~~~?~~~~~   ~~~~~~~~~~~~~~~~~~~~~~~~1971 \pm 15$}\\
{$~~~~~?~~~~~   ~~~~~~~~~~~~~~~~~~~~~~~~2310 \pm 60$}\\

\bigskip
{\bf (1/2,1/2)}\\

{$\omega (0,1^{--})~~~~~~~~~~~~~~~~~~~~~~~b_1(1,1^{+-})$}\\
\medskip
{$1960 \pm 25   ~~~~~~~~~~~~~~~~~~~~~~~~1960 \pm 35$}\\
{$2205 \pm 30   ~~~~~~~~~~~~~~~~~~~~~~~~2240 \pm 35$}\\

\bigskip
{\bf (1/2,1/2)}\\

{$h_1(0,1^{+-})~~~~~~~~~~~~~~~~~~~~~~~\rho(1,1^{--})$}\\
\medskip
{$1965 \pm 45   ~~~~~~~~~~~~~~~~~~~~~~~~1970 \pm 30$}\\
{$2215 \pm 40   ~~~~~~~~~~~~~~~~~~~~~~~~2150 \pm ~?$}\\

\bigskip
{\bf (0,1)+(1,0)}\\

{$a_1(1,1^{++})~~~~~~~~~~~~~~~~~~~~~~~\rho (1,1^{--})$}\\
\medskip
{$1930^{+30}_{-70} ~~~~~~~~~~~~~~~~~~~~~~~~~1900 ~\pm ~?$}\\
{$ 2270^{+55}_{-40} ~~~~~~~~~~~~~~~~~~~~~~~~~2265 \pm 40$}\\

\end{center}
Here, like for the $J=0,2$ states, we again observe  patterns
of chiral symmetry restoration. Two  missing $\omega(0,1^{--})$
states are yet missing.\\

Below are the multiplets for $J=3$.
\begin{center}{\bf (0,0)}\\

{$\omega_3(0,3^{--})~~~~~~~~~~~~~~~~~~~~~~~f_3(0,3^{++})$}\\
\medskip
{$~~~~~?~~~~~   ~~~~~~~~~~~~~~~~~~~~~~~~~2048 \pm 8$}\\
{$2285 \pm 60   ~~~~~~~~~~~~~~~~~~~~~~~~2303 \pm 15$}\\

\bigskip
{\bf (1/2,1/2)}\\

{$\omega_3 (0,3^{--})~~~~~~~~~~~~~~~~~~~~~~~b_3(1,3^{+-})$}\\
\medskip
{$1945 \pm 20   ~~~~~~~~~~~~~~~~~~~~~~~~2032 \pm 12$}\\
{$2255 \pm 15   ~~~~~~~~~~~~~~~~~~~~~~~~2245 \pm ~?$}\\

\bigskip
{\bf (1/2,1/2)}\\

{$h_3(0,3^{+-})~~~~~~~~~~~~~~~~~~~~~~~\rho_3(1,3^{--})$}\\
\medskip
{$2025 \pm 20   ~~~~~~~~~~~~~~~~~~~~~~~~1982 \pm 14$}\\
{$2275 \pm 25   ~~~~~~~~~~~~~~~~~~~~~~~~2260 \pm 20$}\\

\bigskip
{\bf (0,1)+(1,0)}\\

{$a_3(1,3^{++})~~~~~~~~~~~~~~~~~~~~~~~\rho_3 (1,3^{--})$}\\
\medskip
{$2031 \pm 12  ~~~~~~~~~~~~~~~~~~~~~~~~ 2013 \pm 30$}\\
{$2275 \pm 35 ~~~~~~~~~~~~~~~~~~~~~~~~~2300^{~+~50}_{~-~80}$}\\

\end{center}

Data on $J \geq 4$ is scarce.  While some of the multiplets
are well seen, it is not yet possible to provide any
systematic analysis. The prediction is that for $J=4$ the
pattern should be the same as for $J=2$, while for $J=5$ it should
be similar to $J=1,3$ cases.

\section{Evidence for $U(1)_A$ restoration.}

It is important to see whether there are also signatures
of the $U(1)_A$ restoration. This 
can happen if two conditions are fulfilled \cite{CG1}: (i) unimportance
of the axial anomaly in excited states, (ii) chiral
$SU(2)_L \times SU(2)_R$ restoration (i.e. unimportance of the
quark condensates which break simultaneously both types
of symmetries in the vacuum state). 
Some evidence for the $U(1)_A$ restoration
has been reported in ref. \cite{G1} on the basis of $J=0$
data. Yet  missing $a_0$ and $\eta$ states have
to be discovered to complete the $U(1)_A$ multiplets in the
$J=0$ spectra. In this section we will demonstrate that the
data on the $J=1,2,3$ present convincing evidence on $U(1)_A$
restoration.\\

First, we have to consider which mesonic states can be 
expected to be $U(1)_A$ partners. The $U(1)_A$ transformation
connects interpolators of the same isospin but opposite parity.
But not all such interpolators can be connected by the $U(1)_A$
transformation. For instance, the vector currents 
$\bar \psi \gamma^\mu \psi$
and $\bar \psi \vec \tau\gamma^\mu \psi$ are invariant under
$U(1)_A$. Similarly, the axial vector interpolators
 $\bar \psi \gamma^5 \gamma^\mu \psi$
and $\bar \psi \vec \tau \gamma^5 \gamma^\mu \psi$ are also invariant under
$U(1)_A$. Hence those interpolators (states) that are members
of the $(0,0)$ and $(0,1)+(1,0)$ representations of
$SU(2)_L \times SU(2)_R$ are invariant with respect to
$U(1)_A$. However,  interpolators (states) from the {\it distinct}
(1/2,1/2) representations which have the same isospin but
opposite parity  transform into each other under $U(1)_A$.
For example, $\bar \psi \psi \leftrightarrow \bar \psi \gamma^5 \psi$,
 $\bar \psi \vec \tau \psi \leftrightarrow \bar \psi \vec \tau \gamma^5 \psi$,
and those with derivatives: 
$\bar \psi \partial^\mu \psi \leftrightarrow \bar \psi \gamma^5  
\partial^\mu \psi$, $\bar \psi \vec \tau \partial^\mu \psi 
\leftrightarrow \bar \psi \vec \tau \gamma^5  \partial^\mu \psi$, etc.
If the corresponding states are systematically degenerate, then
it is a signal that $U(1)_A$ is restored. In what follows we show
that it is indeed the case.

\begin{center}{\bf J=1}\\

{$\omega(0,1^{--})~~~~~~~~~~~~~~~~~~~~~~~h_1(0,1^{+-})$}\\
\medskip
{$1960 \pm 25   ~~~~~~~~~~~~~~~~~~~~~~~~1965 \pm 45$}\\
{$2205 \pm 30   ~~~~~~~~~~~~~~~~~~~~~~~~2215 \pm 40$}\\

\bigskip

{$b_1 (1,1^{+-})~~~~~~~~~~~~~~~~~~~~~~~\rho(1,1^{--})$}\\
\medskip
{$1960 \pm 35   ~~~~~~~~~~~~~~~~~~~~~~~~1970 \pm 30$}\\
{$2240 \pm 35   ~~~~~~~~~~~~~~~~~~~~~~~~2150 \pm ~?$}\\

\bigskip
{\bf J=2}\\

{$f_2(0,2^{++})~~~~~~~~~~~~~~~~~~~~~~~\eta_2(0,2^{-+})$}\\
\medskip
{$2001 \pm 10  ~~~~~~~~~~~~~~~~~~~~~~~~2030 ~\pm ~?$}\\
{$2293 \pm 13   ~~~~~~~~~~~~~~~~~~~~~~~~2267 \pm 14$}\\

\bigskip

{$\pi_2(1,2^{-+})~~~~~~~~~~~~~~~~~~~~~~~a_2 (1,2^{++})$}\\
\medskip
{$2005 \pm 15  ~~~~~~~~~~~~~~~~~~~~~~~~ 2030 \pm 20$}\\
{$2245 \pm 60 ~~~~~~~~~~~~~~~~~~~~~~~~ 2255 \pm 20 $}\\

\bigskip

{\bf J=3}\\

{$\omega_3(0,3^{--})~~~~~~~~~~~~~~~~~~~~~~~h_3(0,3^{+-})$}\\
\medskip
{$1945 \pm 20   ~~~~~~~~~~~~~~~~~~~~~~~~2025 \pm 20$}\\
{$2255 \pm 15   ~~~~~~~~~~~~~~~~~~~~~~~~2275 \pm 25$}\\

\bigskip

{$b_3 (1,3^{+-})~~~~~~~~~~~~~~~~~~~~~~~\rho_3(1,3^{--})$}\\
\medskip
{$2032 \pm 12   ~~~~~~~~~~~~~~~~~~~~~~~~1982 \pm 14$}\\
{$2245 ~\pm ~?   ~~~~~~~~~~~~~~~~~~~~~~~~2260 \pm 20$}\\

\end{center}

 We see systematic doublets of $U(1)_A$ restoration. Hence
two distinct (1/2,1/2) multiplets of $SU(2)_L \times SU(2)_R$
can be combined into one multiplet of $U(2)_L \times U(2)_R$.
So we conclude that the whole chiral symmetry  of the
QCD Lagrangian $U(2)_L \times U(2)_R$ gets approximately restored high in the hadron spectrum.

\section{Discussion and conclusions}

We have classified $\bar q q$ meson states in the $u,d$
sector according to chiral symmetry of QCD. Then we have presented
recent experimental data on highly excited meson states which
provide evidence that in the high-lying mesons both
$SU(2) \times SU(2)$ and $U(1)_A$ get restored. Hence the
highly excited states should be classified according to
$U(2)_L \times U(2)_R = SU(2)_L \times SU(2)_R \times U(1)_V \times U(1)_A$
symmetry group of QCD Lagrangian. Usual quantum numbers such as
$I,J^{PC}$ are not enough to uniquely specify the states.\\

It is useful to quantify the effect of chiral symmetry breaking
(restoration). An obvious parameter that characterises effects
of chiral symmetry breaking is a relative mass splitting within
the chiral multiplet. Let us define the {\it chiral asymmetry} as

\begin{equation}
\chi = \frac{|M_1 - M_2|}{(M_1+M_2)},
\label{chir}
\end{equation}

\noindent
where $M_1$ and $M_2$ are masses of particles within the
same multiplet. This parameter gives a quantitative measure
of chiral symmetry breaking at the leading (linear) order
and has the interpretation of the part
 of the hadron mass  due to chiral
symmetry breaking.\\

For the low-lying states the chiral asymmetry is typically
0.3 - 0.6 which can be seen e.g. from a comparison of the $\rho(770)$
and $a_1(1260)$ or the  $\rho(770)$ and $h_1(1170)$ masses. If 
the chiral asymmetry is large as above, then it makes no
sense to assign a given hadron to the chiral multiplet
since its wave function is a strong mixture of different
representations and we have to expect also large
{\it nonlinear} symmetry breaking effects. However, at meson 
masses about 2 GeV
the chiral asymmetry is typically within 0.01, as
can be seen from the multiplets presented in this paper,
and in this case the hadrons can be believed to be  members
of  multplets with a tiny admixture of other
representations. Unfortunately there are no systematic data
on mesons below 1.9 GeV and hence it is difficult to estimate
the chiral asymmetry as a function of mass ($\sqrt s$). Such
a function would be crucially important for a further progress
of the theory. So a systematic experimental study of hadron spectra is
difficult to overestimate. However, thanks to the $0^{++}$
glueball search for the last 20 years, there are such data
for $\pi$ and $f_0$ states, as can be seen from Fig. 1
(for details we refer to \cite{G1,G5}). According to these data
we can reconstruct $\chi(\sqrt s \sim 1.3 GeV) \sim 0.03 \div 0.1$,
$\chi(\sqrt s \sim 1.8 GeV) \sim 0.008$, 
$\chi(\sqrt s \sim 2.3 GeV) \sim 0.005$.
We have to also  stress that there is no reason to expect
the chiral asymmetry to be a universal function for all
hadron channels. Hadrons with different quantum numbers
feel chiral symmetry breaking effects differently, as can
be deduced from the operator product expansions of two-point
functions for different currents.
 A task of the theory is to derive these
chiral asymmetries microscopically. A comment on the present theoretical
state of the arts is in order.\\

Naively one would expect that the operator product expansion
of the two-point correlator, which is valid in the deep
Euclidean domain \cite{SVZ}, could help us. This is not
so, however, for two reasons. First of all, we know
phenomenologically only the lowest dimension quark condensate.
Even though this condensate dominates as a chiral symmetry
breaking measure at the very large space-like $Q^2$, at
smaller $Q^2$ the higher dimensional condensates, which
are suppressed by inverse powers of $Q^2$, are also  
important. These condensates are not known, unfortunately.
But even if we knew all  quark condensates
up to a rather high dimension, it would not help us. This is
because the OPE is only an asymptotic expansion \cite{Z}.
While such kind of expansion is very useful in the space-like
region, it does not define any analytical solution which could
be continued to the time-like region at finite $s$. While
convergence of the OPE can be improved by means of the Borel
transform and it makes it useful for SVZ sum rules for the
low-lying hadrons, this cannot be done for the higher states.
So in order to estimate chiral symmetry restoration effects
 one indeed needs a microscopic theory that would incorporate
{\it at the same time} chiral symmetry breaking and confinement.\\

The question  is  which physical picture 
is compatible with this symmetry restoration
for highly excited
mesons. It has been
shown in ref. \cite{G2} that  viewing  excited hadrons  as strings
with quarks of definite chirality at the end points of the string
is consistent with chiral symmetry of QCD. However, we can extend chiral
symmetry to the larger symmetry. If one assumes that the
spin degree of freedom of  quarks (i.e. their helicity) is {\it uncorrelated}
with  the energy of the string, then we have to expect that all
possible states of the string with  all possible helicity
orientations of quarks with the same spin
must have the same energy. This means that
all possible {\it different} $SU(2)_L \times SU(2)_R$  multiplets with the
same $J$ must be degenerate. If we look carefully at the data presented
in this paper we do see  this fact. Does it mean
that we observe a restoration of larger symmetry 
which includes
chiral $U(2)_L \times U(2)_R$ group as a subgroup, i.e.
symmetry which is higher than the symmetry of the QCD Lagrangian !?
There is simpler solution to this problem. There is a {\it reducible}
 representation
which combines (0,0), (1/2,1/2), (1/2,1/2) and (0,1)+(1,0) -
it is $[(0,1/2)\oplus(1/2,0)] \times [(0,1/2)\oplus(1/2,0)]$,
which is still a representation of the parity-chiral group,
i.e. it is still a representation of the symmetry group of the
QCD Lagrangian. This representation is nothing else but a
product of the massless quark and antiquark fields. Hence
a degeneracy within such a reducible representation is indeed
compatible with the string with massless quarks with definite
chirality at the end points of the string.

\section{Acknowledgements}
 
I am grateful to D.V. Bugg for comments on the data \cite{BUGG1,BUGG2,BUGG3,BUGG4}. 
I am also thankful to  C. Lang  for careful reading of the
manuscript.
 The work was supported by the FWF project P16823-N08
of the Austrian Science Fund.

\bigskip
\bigskip
\bigskip

{\bf Table 1} ~Mesons obtained in $\bar p p$ annihilation at 1.9 - 2.4 GeV.
The extra broad $f_2(0,2^{++})$ state with $M = 2010 \pm 25$,
$\Gamma=495 \pm 35$ seen in the $\eta \eta$ channel only is not
included in the Table. If it is a real state, then it can be
either glueball or $\bar s s$ state because it is not seen
in the $\pi \pi$ channel.
\begin{center}
\begin{tabular}{|llllll|} \hline
Meson & ~I~ & $~J^{PC}~$ & Mass (MeV) & Width (MeV) & Reference\\ \hline
$f_1$ & ~0~ & $~1^{++}~$  & $1971 \pm 15$ &  $240 \pm 45$ & \cite{BUGG1}\\
$f_1$ & ~0~ & $~1^{++}~$  & $2310 \pm 60 $ &  $255 \pm 70$ & \cite{BUGG1} \\
$f_2$ & ~0~ & $~2^{++}~$  & $1934 \pm 20$  &  $271 \pm 25$ & \cite{BUGG1} \\
$f_2$ & ~0~ & $~2^{++}~$  & $2001 \pm 10$  &  $312 \pm 32$ & \cite{BUGG1} \\
$f_2$ & ~0~ & $~2^{++}~$  & $2240 \pm 15$  &  $241 \pm 30$ & \cite{BUGG1} \\
$f_2$ & ~0~ & $~2^{++}~$  & $2293 \pm 13$  &  $216 \pm 37$ & \cite{BUGG1} \\
$\eta_2$ & ~0~ & $~2^{-+}~$  & $2030 \pm ? $  &  $205 \pm ?$ & \cite{BUGG1} \\
$\eta_2$ & ~0~ & $~2^{-+}~$  & $2267 \pm 14 $  &  $290 \pm 50$ & \cite{BUGG1} \\
$f_3$ & ~0~ & $~3^{++}~$  & $2048 \pm 8$  &  $213 \pm 34$ & \cite{BUGG1} \\
$f_3$ & ~0~ & $~3^{++}~$  & $2303 \pm 15$  &  $214 \pm 29$ & \cite{BUGG1} \\

$a_1$ & ~1~ & $~1^{++}~$  & $1930^{+30}_{-70} $ &  $155 \pm 45$ & \cite{BUGG2}\\
$a_1$ & ~1~ & $~1^{++}~$  & $2270^{+55}_{-40} $ &  $305^{+70}_{-35}$
   &\cite{BUGG2}\\
$a_2$ & ~1~ & $~2^{++}~$  & $1950^{+30}_{-70} $ &  $180^{+30}_{-70}$
   &\cite{BUGG2}\\
$a_2$ & ~1~ & $~2^{++}~$  & $2030 \pm 20 $ &  $205 \pm 30$ & \cite{BUGG2}\\
$a_2$ & ~1~ & $~2^{++}~$  & $2175 \pm 40 $ &  $310^{+90}_{-45}$ & \cite{BUGG2}\\
$a_2$ & ~1~ & $~2^{++}~$  & $2255 \pm 20 $ &  $230 \pm 15$ & \cite{BUGG2}\\
$\pi_2$ & ~1~ & $~2^{-+}~$  & $2005 \pm 15 $ &  $200 \pm 40$ & \cite{BUGG2}\\
$\pi_2$ & ~1~ & $~2^{-+}~$  & $2245 \pm 60 $ &  $320^{+100}_{-40}$ &  
    \cite{BUGG2}\\
$a_3$ & ~1~ & $~3^{++}~$  & $2031 \pm 12 $ &  $150 \pm 18$ & \cite{BUGG2}\\
$a_3$ & ~1~ & $~3^{++}~$  & $2275 \pm 35 $ &  $350^{+100}_{-50}$ &
 \cite{BUGG2}\\

$\rho$ & ~1~ & $~1^{--}~$  & $1970 \pm 30 $ &  $260 \pm 45$ & \cite{BUGG3}\\
$\rho$ & ~1~ & $~1^{--}~$  & $2110 \pm 35 $ &  $230 \pm 50$ & \cite{BUGG3}\\
$\rho$ & ~1~ & $~1^{--}~$  & $2265 \pm 40 $ &  $325 \pm 80$ & \cite{BUGG3}\\
$b_1$ & ~1~ & $~1^{+-}~$  & $1960 \pm 35 $ &  $230 \pm 50$ & \cite{BUGG3}\\
$b_1$ & ~1~ & $~1^{+-}~$  & $2240 \pm 35 $ &  $320 \pm 85$ & \cite{BUGG3}\\
$\rho_2$ & ~1~ & $~2^{--}~$  & $1940 \pm 40 $ &  $155 \pm 40$ & \cite{BUGG3}\\
$\rho_2$ & ~1~ & $~2^{--}~$  & $2225 \pm 35 $ &  $335^{+100}_{-50}$ &
   \cite{BUGG3}\\
$\rho_3$ & ~1~ & $~3^{--}~$  & $1982 \pm 14 $ &  $188 \pm 24$ & \cite{BUGG3}\\
$\rho_3$ & ~1~ & $~3^{--}~$  & $2013 \pm 30 $ &  $ 165 \pm 35$ & \cite{BUGG5}\\
$\rho_3$ & ~1~ & $~3^{--}~$  & $2260 \pm 20 $ &  $160 \pm 25$ & \cite{BUGG3}\\
$\rho_3$ & ~1~ & $~3^{--}~$  & $2300^{+50}_{-80} $ &  $340 \pm 50$ 
    & \cite{BUGG1}\\
$b_3$ & ~1~ & $~3^{+-}~$  & $2032 \pm 12 $ &  $117 \pm 11$ & \cite{BUGG3}\\
$b_3$ & ~1~ & $~3^{+-}~$  & $2245 \pm ? $ &  $320 \pm 70$ & \cite{BUGG3}\\

$\omega$ & ~0~ & $~1^{--}~$  & $1960 \pm 25 $ &  $195 \pm 60$ & \cite{BUGG4}\\
$\omega$ & ~0~ & $~1^{--}~$  & $2205\pm 30 $ &  $350 \pm 90$ & \cite{BUGG4}\\
$h_1$ & ~0~ & $~1^{+-}~$  & $1965 \pm 45 $ &  $345 \pm 75$ & \cite{BUGG4}\\
$h_1$ & ~0~ & $~1^{+-}~$  & $2215 \pm 40 $ &  $325 \pm 55$ & \cite{BUGG4}\\
$\omega_2$ & ~0~ & $~2^{--}~$  & $1975 \pm 20 $ &  $175 \pm 25$ & \cite{BUGG4}\\
$\omega_2$ & ~0~ & $~2^{--}~$  & $2195 \pm 30 $ &  $225 \pm 40$ & \cite{BUGG4}\\
$\omega_3$ & ~0~ & $~3^{--}~$  & $1945 \pm 20 $ &  $115 \pm 22$ & \cite{BUGG4}\\
$\omega_3$ & ~0~ & $~3^{--}~$  & $2255 \pm 15 $ &  $175 \pm 30$ & \cite{BUGG4}\\
$\omega_3$ & ~0~ & $~3^{--}~$  & $2285 \pm 60 $ &  $230 \pm 40$ & \cite{BUGG4}\\
$h_3$ & ~0~ & $~3^{+-}~$  & $2025 \pm 20 $ &  $145 \pm 30$ & \cite{BUGG4}\\
$h_3$ & ~0~ & $~3^{+-}~$  & $2275 \pm 25 $ &  $190 \pm 45$ & \cite{BUGG4}\\
 \hline
\end{tabular}
\end{center}

\end{document}